\documentclass[11pt]{article} %
\usepackage{fullpage}
\usepackage{graphicx}
\usepackage{graphics}
\usepackage{psfrag}
\usepackage{amsmath, amssymb, graphics, epsfig, enumerate, amstext, amsthm, comment}
\usepackage{hyperref}
\usepackage{xcolor}
\usepackage{amssymb,amsmath}
\usepackage{latexsym,bm,mathrsfs}
\newtheorem{theorem}{Theorem}[section]
\newtheorem{corollary}[theorem]{Corollary}

\newtheorem{example}[theorem]{Example}

\def\la{{\langle}}
\def\ra{{\rangle}}
\newcommand{\bV}{\mathbf{V}}
\newcommand{\bM}{\mathbb{M}}

\newcommand{\IR}{\mathbb{R}}
\newcommand{\IC}{\mathbb{C}}

\def\b0{\mathbf 0}

\def\bu{\mathbf u}

\def\bV{\mathbb V}
\def\cS{\mathcal S}

\def\per{{\rm per}}
\def\ba{{\bf a}}
\def\bb{{\bf b}}
\def\bc{{\bf c}}
\def\bB{{\bf B}}
\begin{document}
\openup 1\jot

\centerline
{\Large\sc A note on Majorana representation of quantum states}

\bigskip
\noindent
Chi-Kwong Li, \\
Department of Mathematics, College of William \& Mary, Williamsburg, VA 23187, USA.
\\
Email: ckli@math.wm.edu

\medskip\noindent
Mikio Nakahara, \\
IQM Quantum Computers, Espoo 02150, Finland.
Email: mikio.nakahara@meetiqm.com\footnote{Mikio Nakahara's
affiliation with IQM Quantum Computers is provided for identification purposes 
only and it is not intended to convey or imply IQM's concurrence with, or support for, the 
positions, opinions, or viewpoints expressed by the author.}

\medskip
\centerline{\bf In memory of Professor
Kalyanapuram Rangachari Parthasarathy.}

\begin{abstract}
By the Majorana representation, for any 
$d > 1$ there is a one-one correspondence between a quantum state of dimension $d$
and $d-1$ qubits represented as $d-1$ points in the Bloch sphere. Using the theory of symmetry class of tensors,
we present a simple scheme for constructing $d-1$ points on the Bloch sphere and the 
corresponding $d-1$ qubits  representing a $d$-dimensional quantum state. 
Additionally, we demonstrate how the inner product of two $d$-dimensional quantum states can 
be expressed as a permanent of a matrix related to their $(d-1)$-qubit state representations. 
Extension of the result to mixed states
is also considered.
\end{abstract}



\section{Introduction}
Quantum states, represented by unit vectors $\mathbf{a}\in\mathbb{C}^d$, form a fundamental aspect of 
$d$-dimensional systems. These vectors are identified up to a phase factor, 
i.e., $\mathbf{a}$ and $e^{it}\mathbf{a}$ are identified 
for any $t \in [0, 2\pi)$. In the case of $d=2$, quantum states are commonly referred to as qubits. 
For qubits, there exists a one-to-one correspondence between the state $\mathbf{a}=(a_0, a_1)^t$ with $|a_0|^2 + |a_1|^2 = 1$
and a point $(c_x, c_y, c_z)$ on the Bloch sphere:
$$\bB = \{ (c_x,c_y,c_z):
c_x, c_y, c_z \in \IR^3, \ c_x^2+c_y^2 + c_z^2 = 1\}$$ 
where $\mathbf{a}\mathbf{a}^*$ corresponds to $(c_x, c_y, c_z)$ by 
$\frac{1}{2}\begin{pmatrix}1 + c_z & c_x-ic_y\cr c_x+ic_y & 1-c_z\cr\end{pmatrix}$. The 
correspondence is established using 
$(c_x, c_y, c_z) = (\Re(\bar a_0 a_1), \Im(\bar a_0 a_1), |a_0|^2-|a_1|^2)/2$, ensuring that 
$c_x^2+ c_y^2 + c_z^2 = 1$.

In \cite{Majorana}, Majorana proposed a geometric method to represent a quantum state 
$\mathbf{a} \in \mathbb{C}^d$ for $d>1$ using $d-1$ qubits. Consequently, a quantum state in $
\mathbb{C}^d$ is associated with $d-1$ points on the Bloch sphere. 
The Majorana representation provides a visual tool to understand the properties and 
transformations of quantum states. The direct visualization of qubit rotations are 
useful in the study of different topics of quantum information science such as quantum computation 
and communication; e.g., see \cite{SB} and its references.

In this note, we establish a connection between the Majorana representation and the symmetry class 
of  tensors 
in $\mathbb{V}^{\otimes (d-1)}$ for $\mathbb{V} = \mathbb{C}^2$ associated with the principal 
character $\xi$. 
Using this connection, we  provide an easy scheme to determine 
$v_1, \dots, v_{d-1} \in \mathbb{C}^2$ associated with a given vector
$(a_0, \dots, a_{d-1}) \in \IC^d$.
Additionally, we present a simple formula for the inner product of $\mathbf{a}$ and 
$\mathbf{b}$ in $\mathbb{C}^d$ in terms of their $(d-1)$-qubit presentations. Numerical 
examples are given to illustrate the result. Extension of the result to mixed states is also considered.

\section{Results}

\subsection{Preliminary}

Let us present the following standard set up of 
a symmetry class of tensors in the 
$(d-1)$-fold tensor space $\bV^{\otimes (d-1)}$.
In our study
we focus on $\bV = \IC^2$ and the principal character $\xi$ on the 
symmetric group $S_{d-1}$ of degree $d-1$ such that 
$\xi(\sigma) =1$ for all $\sigma \in S_{d-1}$.
Define the symmetrizer on the tensor space $\bV^{\otimes (d-1)}$ by
$$ T(v_1 \otimes \cdots \otimes v_{d-1}) 
= \frac{1}{(d-1)!} \sum_{ \sigma\in S_{d-1}}
\xi(\sigma)  v_{\sigma^{-1}(1)} \otimes \cdots \otimes v_{\sigma^{-1}(d-1)}.$$
Then 
$\bV_\xi^{(d-1)} = T(\bV^{\otimes (d-1)})$
is a subspace of $\bV^{\otimes (d-1)}$ known as the 
symmetry class of tensors over $\bV$ associated
with $\xi$ on $S_{d-1}$. The elements in  $\bV_\xi^{(d-1)}$ of the form
$T(v_1 \otimes \cdots \otimes v_{d-1})$ are
called decomposable tensors and are denoted by $v^\bullet = v_1 \bullet \cdots \bullet  v_{d-1}$.
One may  see \cite{Grueb,Marcus} for some 
general background. In fact, researchers have used decomposable tensors to model 
boson states; see \cite{BLP}.

Let $\{e_0, e_1\}$ be the standard orthonormal basis of $\bV = \IC^2$ using the standard
inner product $\la u, v\ra = u^*v$, where $X^*$  denotes the 
conjugate transpose of $X$ if $X$ is a complex vector or matrix. 
Then 
$\bV_\xi^{(d-1)}$ is the subspace of $\bV^{\otimes (d-1)}$ spanned by the orthogonal basis
$$\cS = \{e_{i_1} \bullet \cdots \bullet e_{i_{d-1}}: 
0 \le i_1 \le \cdots \le i_{d-1} \le 1\}$$
using the induced inner product on decomposable tensor 
$u_1 \bullet\cdots \bullet u_{d-1}$
and $v_1 \bullet \cdots \bullet v_{d-1}$ so that
$$\la u_1 \bullet\cdots \bullet u_{d-1}, v_1 \bullet \cdots \bullet v_{d-1}\ra = 
\frac{1}{(d-1)!}\per(\la u_i,v_j\ra),$$
where 
$$\per(X) = \sum_{\sigma \in S_k}\prod_{j=1}^k 
x_{j\sigma(j)} \quad \hbox{ for } X = (x_{ij}) \in \bM_k$$
is the permanent of $X \in \bM_k$; see e.g., \cite{Minc} for basic properties of the permanent.
If $j_1 = \cdots = j_\ell = 0$
and $j_{\ell+1} = \cdots = j_{d-1} = 1$, then 
$$\la e_{j_1} \bullet \cdots \bullet e_{j_{d-1}}, e_{j_1} \bullet \cdots \bullet e_{j_{d-1}}  \ra = 
\per\left(J_\ell \oplus J_{d-1-\ell} \right)/(d-1)! = {d-1\choose \ell}^{-1},$$
where $J_r \in M_r$ has all entries equal to 1.
Thus, after normalization $\cS$ becomes an orthonormal basis
$\{f_0^\bullet, \dots, f_{d-1}^\bullet\}$.
Let
\begin{equation}\label{def-C}
C_j = [\underbrace{e_0 \ \cdots \ e_0}_{d-1-j} \ \underbrace{e_1 \ \cdots \ e_1}_{j}]\in \bM_{2,d-1}, \qquad j = 0, \dots, d-1.
\end{equation} 
Suppose  $v_1, \dots, v_{d-1}\in \IC^2$. Then the 
decomposable tensor 
\begin{equation} \label{eq}
v^\bullet = v_1\bullet\cdots \bullet v_{d-1} 
= a_0 f_0^\bullet + \cdots + a_{d-1} f_{d-1}^\bullet
\end{equation}
with 
\begin{equation}
\label{aj}
a_j = \la f_j^{\bullet},v^\bullet\ra =  \frac{1}{(d-1)!}
\sqrt{d-1\choose j}\per(C_j^* [v_1 \cdots v_{d-1}]), \qquad j = 0, \dots, d-1.
\end{equation}
Moreover,
\begin{itemize}
\item[(a)] $\gamma(v_1\bullet \cdots \bullet v_{d-1}) 
= \mu_1 v_1\bullet \cdots \bullet \mu_{d-1} v_{d-1}$ if 
$\mu_1, \dots, \mu_{d-1}, \gamma \in \IC$ satisfy $\mu_1 \cdots \mu_{d-1} = \gamma$, 

\item[(b)] $v_1\bullet \cdots \bullet v_{d-1} =  v_{\sigma(1)}\bullet \cdots \bullet 
v_{\sigma(d-1)}$
if $\sigma \in S_{d-1}$ is a permutation of $(1, \dots, d-1)$.
\end{itemize}

\medskip
We will also use the following fact about the zeros of a complex polynomial.
Let $E_k(\mu_1,\dots, \mu_{d-1})$ be the 
$k$th elementary symmetric function for $\mu_1, \dots, \mu_{d-1}$, i.e., 
$$ E_k(\mu_1, \dots, \mu_{d-1}) = \sum_{1 \le j_1 < \cdots < j_k
\le d-1} \mu_{j_1} \cdots \mu_{j_k},  \qquad k = 1, \dots, d-1.$$ 
Let 
\begin{eqnarray*}
g(z) &=& c_0z^{d-1} + c_1 z_{d-2} + \cdots + c_{d-1} 
= c_0(z^{d-1} + \frac{c_1}{c_0} z_{d-2} + \cdots + \frac{c_{d-1}}{c_0}) \\ 
&=& c_0(z-\mu_1)\cdots (z-\mu_{d-1}),
\end{eqnarray*}
where $c_0\ne 0.$ 
Then 
$$\frac{c_j}{c_0} = (-1)^j E_j(\mu_1, \dots, \mu_{d-1}), \qquad j = 1, \dots, d-1.$$

\subsection{Main result and examples}

\begin{theorem} \label{main} Let $\bV = \IC^2$ and 
$\{f_0^\bullet, \dots, f_{d-1}^\bullet\}$ be the standard orthonormal
basis for $\bV_\xi^{(d-1)}$.  If $(a_0, \dots, a_{d-1})^t \in \IC^{d}$ is nonzero 
and $r\ge 0$ is the smallest integer such that $a_r \ne 0$, then 
for $\gamma_r = a_r\sqrt{{d-1\choose r}}$
$$\frac{1}{\gamma_r}(a_0f_0^\bullet + \cdots + a_{d-1} f_{d-1}^\bullet) = v_1\bullet \cdots \bullet v_{d-1}$$
so that 
$$
a_0f_0^\bullet + \cdots + a_{d-1} f_{d-1}^\bullet = v_1\bullet \cdots \bullet v_{d-2} \bullet (\gamma_r v_{d-1})
$$ 
with 
$v_1 = \cdots = v_{r} = (0,1)^t$, and 
$v_j  = (1, \mu_j)^t$ for $j = r+1, \dots, d-1,$
where 
$\mu_{r+1}, \dots, \mu_{d-1}$ are the zeros of the Majorana polynomial
$$g(z) = \sum_{j=0}^{d-1} (-1)^{j} a_{j}\sqrt{d-1\choose j}z^{d-1-j}.$$ 
 If
$b_0 f_0^\bullet + \cdots + b_{d-1} f_{d-1}^\bullet = u_1 \bullet \cdots \bullet u_{d-1}$ and
$c_0 f_0^\bullet + \cdots + c_{d-1} f_{d-1}^\bullet = w_1\bullet \cdots \bullet w_{d-1}$, then
$$\sum_{j=0}^{d-1} \bar b_j c_j = 
\la u_1\bullet \cdots \bullet u_{d-1}, w_1\bullet \cdots \bullet w_{d-1}\ra 
= \per(\la u_i,w_j\ra)/(d-1)!.$$
\end{theorem}

By Theorem \ref{main}, every vector $f^\bullet \in  \bV_\xi^{(d-1)}$ 
admits a representation of the form $u_1\bullet \cdots \bullet u_{d-1}$.
In particular, if $(a_0, \dots, a_{d-1})^t \in \IC^d$ is a quantum state, i.e., a unit vector, then 
$$a_0 f_0^\bullet + \cdots + a_{d-1}f_{d-1}^\bullet 
=\frac{\gamma_r}{|\gamma_r| \gamma }\left(\frac{v_1}{\|v_1\|} \bullet \cdots \bullet \frac{v_{d-1}}{\|v_{d-1}\|}\right),$$
where $\gamma_r, v_1, \dots, v_{d-1}$ are defined as in Theorem \ref{main} and 
$$\gamma = \| \left( \frac{v_1}{\|v_1\|} \bullet \cdots \bullet \frac{v_{d-1}}{\|v_{d-1}\|} \right) \|
=   \prod_{j=1}^{d-1} \|v_j\|\left\{\frac{1}{(d-1)!}\per(\la v_i, v_j\ra)\right\}^{1/2} .$$

\medskip\noindent
\it Proof of Theorem \ref{main}.
\rm 
If $r = d-1$,  then clearly $\frac{1}{\gamma_r}(a_r f^\bullet_{d-1}) =
\frac{1}{a_r}(a_r f^\bullet_{d-1}) =  e_1 \bullet \cdots \bullet e_1$.

\medskip
Suppose $r < d-1$.  Construct the vectors $v_1, \dots, v_{d-1}$ as described.
We will show that
$$\frac{1}{a_r\sqrt{{d-1\choose r}}}(a_0 f_0^\bullet + \cdots a_{d-1} f_{d-1}^\bullet)  =  v_1 \bullet \cdots  \bullet  v_{d-1}.$$
Let ${\bf 1}_k \in \IC^k$ has all entries equal to 1,
$C_j$ be defined as in (\ref{def-C}), and 
$Q\in \bM_{2,d-1}$ have columns
$v_1, \dots, v_{d-1}$.
Then 
$$C_j^* Q =  \begin{pmatrix} 0_{d-1-j,r-1} & {\bf 1}_{d-1-j} {\bf 1}_{d-r}^t\cr
{\bf 1}_{j} {\bf 1}_{r-1}^t & {\bf 1}_{j} (\mu_{r}, \dots, \mu_{d-1})\cr
\end{pmatrix}.
$$
By a direct computation, say, using the Laplace expansion formula for permanent and induction, 
we have the following. 
For $j = 0, \dots, r-1$, we have
$\per(C_j^* Q) = 0$ and hence 
$$\la f_j^{\bullet}, v_1\bullet\dots \bullet v_{d-1}\ra = 
\sqrt{d-1\choose j} \per(C_j^* Q)/(d-1)! = 0.$$
For $j = r, \dots, d-1$, we have
$\per(C_j^* Q) =   j!(d-1-j)!
E_{j-r}(\mu_{r+1}, \dots, \mu_{d-1})$, 
and hence
\begin{eqnarray*}
\la f_j^{\bullet}, v_1\bullet\cdots \bullet v_{d-1}\ra &=& 
\sqrt{d-1\choose j} \per(C_j^* Q)/(d-1)! \\
& = &
E_{j-r}(\mu_{r+1}, \dots, \mu_{d-1})/\sqrt{d-1 \choose j}.
\end{eqnarray*}
Since $\mu_r, \dots, \mu_{d-1}$ are the zeros of $g(x)$, 
$$E_{j-r}(\mu_{r+1}, \dots, \mu_{d-1})/ \sqrt{{d-1\choose j}} = \frac{a_j}{a_r\sqrt{{d-1\choose r}}} = \frac{a_j}{\gamma_r}.$$
Hence,
$$\frac{1}{\gamma_r}(a_0f_0^\bullet + \cdots + a_{d-1} f_{d-1}^\bullet) = v_1 \bullet \cdots \bullet v_{d-1}.$$ 
The last statement is clear. \qed

\medskip
The following numerical examples illustrate Theorem \ref{main}.

\begin{example} \rm
Suppose $d = 5$ and $\ba  = (a_0,a_1,a_2,a_3,a_4)^t \in \IC^5$ be a nonzero vector.
Let $r\ge 0$ be the smallest integer such that $a_r \ne 0$.  Then for $\gamma_r = a_r\sqrt{4\choose r}$,
$$\frac{1}{\gamma_r}(a_0f_0^\bullet + a_1 f_1^\bullet + a_2f_2^\bullet + a_3 f_3^\bullet +a_4f_4^\bullet)= 
v_1\bullet v_2\bullet v_3\bullet v_4,$$
with  $v_1 = \cdots = v_r = (0,1)^t$ and
$v_{j} = (1, \mu_{j})^t$ for $j = r+1, \dots, 4$, where $\mu_{r+1}, \dots, \mu_4$
are the zeros of the Majorana polynomial
$$g(z) = \sqrt{{4\choose 0}}a_0z^4 -  \sqrt{{4\choose 1}} a_2z^3 + 
 \sqrt{{4\choose 2}} a_2z^2 -  \sqrt{{4\choose 3}}a_3z +  \sqrt{{4\choose 4}}a_4.$$
\end{example}

\begin{itemize}
\item[(i)] Let $\ba =(1,3, 13/\sqrt 6, 6, 4)^t \in \IC^5$.
Then $g(z) = z^4 - 6z^3 + 13z^2 - 12z + 4  = (z-1)^2(z-2)^2$ so that
$\ba$ corresponds to 
$u_1 \bullet u_2 \bullet u_3 \bullet u_4$ with $u_1 = u_2 = (1,1)^t$ and $u_3 = u_4 = (1,2)^t$.

\item[(ii)] Let
$\bb =(0, 1/2, \sqrt 6, 11/2, 6)^t$.
Then $g(z)  = z^3 - 6z^2 +11 z - 6 = (z-1)(z-2)(z-3)$ so that 
$\bb$ corresponds to 
$v_1\bullet \cdots \bullet v_4$ with $v_1 = (0,1)^t$
and $v_2  = (1,1)^t$, $v_3 = (1,2)^t$ and $v_4 = (1,3)^t$.

\item[(iii)]
Let $\bc = (0,0,1/\sqrt 6, 1, 1)^t$.
Then $g(z) = z^2-2z+1 = (z-1)^2$ so that 
$\bc$ corresponds to 
$w_1\bullet \cdots \bullet w_4$ with 
$w_1 = w_2 = (0,1)^t$, and $w_3 = w_4 = (1,1)^t$.

\end{itemize}
\noindent
We have \hskip .63in
$\la \ba, \bb\ra = \per([u_1 u_2 u_3 u_4]^*[v_1 v_2 v_3 v_4])/4! = 143/2,$

\medskip\hskip 1in
$\la \ba, \bc \ra = \per([u_1 u_2 u_3 u_4]^*[w_1 w_2 w_3 w_4])/4! = 12+1/6,$
and 

\medskip\hskip 1in 
$\la \bb, \bc\ra = 
 \per([v_1 v_2 v_3 v_4]^*[w_1 w_2 w_3 w_4])/4! = 25/2.$

\subsection{Mixed states}

Recall that a general quantum state is called a mixed state and is
represented by a density matrix $\rho$, which is 
a positive semi-definite matrix with trace 1. If $\rho$ is rank one, i.e., $\rho = \ba \ba^*$ for a unit vector
 $\ba = (a_0, \dots, a_{d-1})^t \in \IC^d$, then $\rho$ is pure state. If 
 $\ba$ corresponds to $u_1 \bullet \cdots \bullet u_{d-1} \in \bV_\xi^{d-1}$, then by (\ref{aj})
the corresponding density matrix 
$\rho = \ba \ba^* \in \bM_{d}$ has $(r,s)$ entry
equal to 
$$a_r\bar a_s  = \sqrt{{d-1\choose r}{d-1\choose s}}
\frac{\per(C_r^* [u_1 \cdots u_{d-1}])
\per([u_1 \cdots u_{d-1}]^* C_s])}{(d-1)!(d-1)!}, \qquad
0 \le r, s \le d-1,$$
where $C_j$ is defined as in {\rm (\ref{def-C})}. 

There has been interest in finding the Majorana representation for mixed states; e.g., see \cite{SB}.
A general mixed state can be written as
$\rho = \sum_{j=1}^r p_j  \rho_j\in \bM_d$, where $(p_1, \dots, p_r)$ is a probability
vector and $\rho_1,\dots, \rho_r$ are pure states. We can apply Theorem \ref{main}
to each $\rho_j$,  and express it in terms of 
$d-1$ qubit states. Then the mixed state $\rho$ can be associated with a collection of $r$ sets of 
qubit states each has $d-1$ elements and a probability vector $(p_1, \dots, p_r)$.

Alternatively, by purification one may express $\rho$ as the partial trace
of a pure stat $|\psi\ra \la \psi|$ with $|\psi\ra \in \IC^{rd}$, which admits a Majorana representation of 
$rd-1$ qubits.

\bigskip\noindent
{\bf \large Acknowledgment}

\smallskip
The authors would like to thank
Karol \.{Z}yczkowski, Marcin Rudzi{\'n}ski, and the referee for some helpful comments. 
Li is an affiliate member of the Institute for Quantum Computing, University of Waterloo.
His research was supported by the Simons Foundation Grant 851334.

\end{document}